\documentstyle[12pt]{article}

\begin{document}
\baselineskip=18.6pt plus 0.2pt minus 0.1pt \makeatletter
\@addtoreset{equation}{section} \renewcommand{\theequation}{\thesection.%
\arabic{equation}} \makeatletter \@addtoreset{equation}{section}

\begin{titlepage}

\title{\vspace{-3cm}
\hfill\parbox{4cm}{\normalsize LPHEA /00-06}\\
\hfill\parbox{4cm}{\normalsize UFR-HEP 00/24}\\
\hfill\parbox{4cm}{\normalsize November 2000}\\
 \vspace{1cm}
 On the Dilaton and the Axion  Potentials}
 \vspace{2cm}

\author{M. Chabab$^{1.2}$\thanks{mchabab@ucam.ac.ma},
 N. El Biaze$^{1}$, R. Markazi$^1$ and E.H. Saidi$^1$\\
\
\small{$^1$ Lab/UFR High Energy Physics, Faculty of Science, P. O.
Box 1014 , Rabat, Morocco.}\\ \small{$^2$ LPHEA, Physics
Department, Faculty of  Science-Semlalia, P. O. Box 2390,
Marrakesh, Morocco.}} \maketitle \setcounter{page}{1}

\begin{abstract}
We extend the Vecchia-Veneziano-Witten (VVW) model of QCD in the
chiral limit and for large colour number $N_c$, by introducing an
effective dilaton-gluon coupling from which we derive both the
axion and dilaton potentials. Furthermore, using a string inspired
model, we determine a new interquark potential as a perturbative
series in terms of the interquark distance $r$. Our potential goes
beyond Dick one obtained in \cite{D} and shares the same features
as the Bian-Huang-Shen potential $V_{BHS}$  which depends only on
odd powers of $r$ \cite{BHS}.
\end{abstract}
\end{titlepage}

\newpage
\section{Introduction}
\qquad The axion $a$ and the dilaton $\phi$ are two particles
predicted by superstring theories. They are among the probable
candidates which could provide more informations about the Quantum
Chromodynamics (QCD) vacuum \cite{ T, LR, D1}. The latter has a
non trivial structure manifested by the presence of non-zero
vacuum expectation values of certain operators. Theoretically, the
axion was firstly introduced in QCD via the Peccei-Quinn mechanism
where it appears as a pseudo-Goldstone boson resulting from the
spontaneous breakdown of the $ U(1)_{PQ}$ global symmetry in order
to solve the strong CP problem \cite{PQ}. This phenomenon is
directly related to the presence of the topological term $ \theta
{g^2 \over 32 \pi }G_{\mu  \nu} \tilde G^{\mu \nu} $ violating
both P and CP symmetries \cite{PQ} and playing a crucial role in
the Seiberg-Witten model \cite{SW}. Here $ G_{\mu \nu} $ is the
QCD field strength, $ \tilde G_{\mu  \nu}$ its dual and $\theta $
denotes the usual QCD vacuum angle. Furthermore, in cosmology, the
axion is an hypothetical particle presented as a dark matter
candidate and may constitute the missing mass of universe
\cite{S}. As to the dilaton, it is a scalar field appearing in
several theories with different meanings, sometimes as a massless
field and others as a massive one. In supersymmetric gauge
theories, the dilaton $\phi$ appears as a companion of the axion:
if the dilaton arises in the real part of the lowest component of
a chiral superfield, the imaginary part corresponds to an axionic
field. Within the Kaluza-Klein supergravity frame, if the field
content of space-time is assumed to arise from embedding in
$(4+d_i)$-dimensional manifold, the $d_i$ internal dimensions will
have variable volume from the four-dimensional point of view and
is interpreted as the dilaton. In string theory, the axion and the
dilaton are in the masseless spectrum of $10$D type IIB model and
its lower dimensional compactifications. The field $\phi$ emerges
from the NS-NS sector and $a$ from the RR sector and are at the
basis of the discovery of F-theory and type IIB self duality \cite
{VA}. Thus, there is no unique definition of the dilaton as it has
various interacting ways; for instance, it couples naturally to
super-Yang Mills gauge fields in curved space and plays a central
role in string theory since it defines the string coupling
constant $g_s$ as $e^{\phi}$. The dilaton has also been coupled to
the Ricci curvature term as in the Brans-Dicke model of induced
gravity as well as to the QCD field strength to generate the
confining phase. In this case, it was shown in \cite{D} that the
coupling of a massive dilaton to gluon can generate a potential,
of the quark-quark interactions, which exhibits a short range
Coulomb potential plus a linear confining term at large distances.
In this paper, we shall determine the QCD nonperturbative effects,
involved in the axion potential and the interquark potential by
the inclusion of dilatonic degrees of freedom. Specifically, we
shall extend the Vecchia, Veneziano and Witten (VVW) \cite{VV,W}
model, of standard QCD in the chiral limit and for large colour
number $N_c$, by introducing a string inspired effective
dilaton-gluon coupling from which we shall derive both the axion
and dilaton potentials. On the other hand, following \cite{D}, we
shall re-analyze the Coulomb problem under the requirement that
analytical solutions exist, other than Dick one. By making a
simple assumption, a new solution for the dilaton-gluon coupling
function will be determined, from which we shall generate an
interquark potential which goes beyond Dick one and looks like the
Bian-Huang-Shen potential \cite{BHS}. The latter includes
corrections from nonperturbative QCD, encoded in the quarks and
gluons vacuum condensates, and depends only on odd powers of
interquark distance r.
\\

\qquad This paper is organized as follows: In Section 2, first we
review the axion potential, using the gluodynamics effective model
of Halperin and Zhitnitsky \cite{HZ}. Then, we derive new
solutions for the potential. In section 3, we study an extension
of the VVW potential by considering the dilatonic effects. Section
4 is devoted to derive a new interquark potential with higher
orders in $r$ using a massive dilaton field which couples to the
YM kinetic term like ${1\over F(\phi )}G_{\mu \nu}G^{\mu \nu}$. In
section 5, we draw a general conclusion.

\section{Comments on the axion potential}

\qquad We start by recalling that the vacuum energy $E(\theta )$
can be expressed through the  $\theta$-angle appearing  in the CP
violating term of QCD lagrangian  as follows \cite{G1}:
\begin {equation}
E( \theta )=E(0)-{1 \over 2}\chi  \theta ^2,
\end {equation}
where $\chi $ is the topological susceptibility given in terms of
the light quark mass $m_q$ via the following formula \cite{SVZ1}
:\\
\begin {equation}
\chi= i\int dx <0|T \{ {\alpha_s \over 8 \pi} G \tilde
G(x),{\alpha _s \over 8 \pi} G \tilde G(0)\}|0>=
 {m_q\over N_f}<0|\bar \psi \psi |0>+O(m_{q}^2),
\end {equation}
with $N_f$ is the number of the flavors and $<0|\bar \psi \psi
|0>$ is the quark condensate.\\
 Following \cite{HZ}, the one-to-one correspondence $
\theta \to {a \over {f_a}}$   allows us to derive the axion
potential V(a) from Eq.(2.1),
\begin {equation}
V(a) = V(0)-{1 \over 2}\chi ({ a \over f_a })^2
\end {equation}
where $f_a$ denotes the axion coupling constant. The equation
(2.1) appears insufficient to describe interesting properties as
it is not a periodic function of $\theta$. $\theta$ periodicity is
an essential property to investigate vacuum features, especially
for the derivation of the metastable states  and the domain walls
\cite{GAG}.\\ A way to exhibit this periodicity in $ \theta $ is
to consider the Vecchia-Venezino-Witten (VVW) model of standard
QCD. Its lagrangian in the chiral limit and for large colour
number $N_c$ \cite {VV, W} is given by,
\begin {equation}
 L= {F_\pi \over 2}\left( Tr (\partial _\mu U  \partial ^\mu U^+) + Tr(MU+M^+U ^+)\right)
 +{c \over N_c}\left(-i\ log (det U)- \theta\right)^2
\end {equation}
where $M$ is the mass matrix. By minimizing the potential of this
model and taking the diagonal parameterization of the linear
$\sigma$ model U matrix,
\begin{equation}
U=\left( \matrix { &e^{-i\phi_1}   &0         &0 \cr &0
&e^{-i\phi_2}   &0\cr &0            &0            &e^{-i\phi_3}&
\cr}\right)
\end{equation}
one find out that: (i) the obtained minima is invariant by shifts
of $\theta $ as $ \theta +2 \pi $; (ii) Physical observables are
$\theta $ periodic  \cite {VV,W,SM}.\\ An other way to derive the
axion potential relies on the use of the Halperin-Zhitnitsky (HZ)
model. In this model, one proceeds in two steps: the first step
consists in  building the gluodynamics potential while in the
second step one uses the results of Veneziano-Yankielowicz (VY)
supersymmetric model \cite{VY} as well as VVW model.\\
 \qquad The  gluodynamics potential is
expressed as a Legendre transformation in terms of the sources $J$
and $\bar J $ and the so-called  " glueball" fields  h and $\bar
h$ via the following  formula \cite{HZ, GS}:
\begin{equation}
-V\hat U(h, \bar h)=W(h, \bar h)+J \int dx h+\bar J \int dx \bar h
\end{equation}
where $V=\int d^4x$  is  four dimensional volume, $\hat U(h, \bar
h)$ defines the glueballs potential and $W(h, \bar h )$ is
inherited from the action of the gluodynamics lagrangian
consisting only of YM term $G_{\mu \nu }G^{\mu \nu }$ and the
topological term $ \theta {g^2 \over 32 \pi }G_{\mu \nu } \tilde
G^{\mu \nu }$. It obeys the following differential equations:
\begin{eqnarray}
&&{\partial ^{n+1} W \over \partial J ^{n+1}}|_{J=\bar J=0}=(-4)^n
\int dx <H> \nonumber \\ &&{\partial ^{n+1} W \over \partial \bar
J ^{n+1}}|_{J=\bar J=0}=(-4)^n\int dx <\bar H> \\ &&{\partial
^{l+k} W \over \partial J ^l\partial \bar J ^{k}}|_{J=\bar J=0}=0,
\nonumber
\end{eqnarray}
which are simply a generalization of the low energy theorems
written as two points functions (n=1):
\begin{eqnarray}
&&lim_{q \to 0}i\int dx e^{iqx}<0|T \{ H(x)H(0) \}
|0>=-4<H>\nonumber \\ &&lim_{q \to 0}i\int dx e^{iqx}<0|T \{ \bar
H(x)\bar H(0) \} |0>=-4< \bar  H> \\ &&lim_{q \to 0}i\int dx
e^{iqx}<0|T \{ \bar H(x)H(0)\} |0>=0, \nonumber
\end {eqnarray}
with  $ H=(G^2+{i \over \xi}G\tilde G )$ and $ \bar H=(G^2-{i
\over \xi}G\tilde G)$ and where $\xi$ is a parameter related to
the topological susceptibility \cite{HZ}. Moreover, $W(h, \bar h)$
can be expressed in terms of the glueball fields $h$ and $\bar h$
via the following relations:
\begin{eqnarray}
&&{\partial W \over \partial J}=\int dx h \nonumber \\ &&{\partial
W \over \partial \bar J}=\int dx \bar h.
\end{eqnarray}
In order to give an explicit formula of the gluodynamics effective
potential ${\hat U}(h, \bar h)$, it is necessary to derive the
expression of $W(h, \bar h)$. Following \cite{HZ}, the following
form given by:\\
\begin{equation}
W(h,\bar h)=-{1 \over 4} \int dx < H>e^{-4 J}-{1 \over 4} \int dx
<\bar H>e^{-4\bar J},
\end{equation}
 satisfies (2.7) and leads to the effective potential,
\\
\begin{equation}
 \hat U(h,\bar h)={1 \over 4}\,{q \over p}\, h\, log {({h \over C})^ {p \over q}}+{1 \over 4}
 \,{q \over p}\,\bar h \, log {({\bar h \over \bar C})}^ {p \over
q}+D(h-\bar h),
\end{equation}
where the matrix elements  $<H>$, $<\bar H>$ and the $\theta $
vacuum angle are encoded in $C, \bar C, D$ and in the glueball
fields $h$ and $ \bar h$. \\

Now we start the second step to derive  the axion potential: We
consider an effective potential inspired from the VY
supersymmetric model which combines the gluodynamics potential
(2.11) with the linear $\sigma$ model matrix U defined in (2.5):
\begin{eqnarray}
e^{-iVW(h,U)} &=&\Sigma _{n}\Sigma _{k}exp\left\{ -{\frac{i V}{4}}\left( {%
\frac{p}{q}}\,h \,log \left(
({\frac{h}{2eE}})^{\frac{p}{q}}\,detU\right) -2Tr(M U)+h.c.\right)
\right.   \nonumber \\
&&\left. +i\pi V(k+{\frac{q}{p}}\,{\frac{\theta +2\pi n}{2\pi }}){\frac{(h-%
\bar{h})}{2i}}\right\}
\end{eqnarray}
with $ E=b<{\alpha _s \over {32 \pi}}G^2>$, and where $b$ is the
first coefficient of the $\beta $-function and $p$ and $q$ are
integers with $q/p \sim N_c$. Then, once the minimization
procedure is performed, we obtain the following potential:
\begin{equation}
W_{QCD}( \theta,U,U^+)= E cos\left( -\frac{p}{q} \left( \theta
-i\log \left( \det U\right) \right) \right) +{1 \over 2}
Tr(MU+M^+U^+).
\end{equation}
where  $ \rho $ and $\eta $ are physical fields appearing in the
glueball field as $h=E e^{\rho+i \eta}$.\\
 Note that the expansion of the cosine to the next-to-leading order in $ \theta $ produces
a part of the potential similar to VVW one. The latter is derived
in the chiral limit and for large value of colour number $N_c$. By
considering the missing terms at order up to ${1 \over N_c^2}$, we
obtain the generalized formula of the vacuum energy:
\begin{equation}
E(\theta )= E + \chi \, ({q\over p})^2 cos({p \over q}\theta).
\end{equation}
Finally, by invoking again the one-to-one correspondence: $\theta
\to {a \over {f_a}}$ and $E(\theta ) \to V({a \over {f_a}})$, the
axion potential is derived,
\\
\begin{equation}
V(a )=-b<0|{\alpha  _s  \over {32 \pi }}G^2|0>\, + \,m_q N_c
<0|\bar \psi \psi  |0> cos ({a \over {f_a N_c}}).
\end{equation}
This expression exhibits an explicit $\theta$ periodicity as
required and has the same form as a class of inflationary
potential describing the vacuum \cite{KM}.
  At this stage, a couple of remarks are in
order:\\
(i) we can obtain the same result as in (2.13) by an
appropriate choice of $h$ and $\bar h$ incorporated in the
following potential:
\begin{equation}
\hat U(h,\bar h)={1 \over 4}\,h \,log {h \over C ' } + {1 \over
4}\,\bar h \,log {\bar h \over C' }-i\ D'  (h-\bar h),
\end{equation}
where $C' $ and $D ' $ are real. Indeed by taking $ h=E e^{iD' }$,
a straightforward calculation provides the expression (2.13), with
$C' ={E \over e}$ and $4D' ={p \over q} ( \theta -i\ log \,(det
U))$ and where the glueball fields $h$ and $ \bar h$ are
parameterized via the $\theta$-angle and the $U$-matrix.\\ (ii) we
also note that the system of differential equations (2.7) admits
other solutions which are different from the HZ one (2.10).
Consequently, other forms for $\hat U(h, \bar h)$ can be obtained.
In this paper, we propose the following alternative ones:\\
 a)The first solution corresponds to:
\begin{equation}
W(h,\bar h)=-{1 \over {4(n+1)!}}{(1-4J)^{n+1}} \int dx < H>-{1
\over {4(n+1)!}} {(1-4 \bar J)^{n+1}} \int dx <\bar  H>.
\end{equation}

By putting back (2.17) in (2.6) and taking into account the
constraints (2.7) and (2.9),  the following equation emerges:
\begin{equation}
\hat U-h{\partial \hat U \over \partial h}-\bar h{\partial \hat U
\over \partial \bar h}={1 \over 8}{h \over  < H>}+{1 \over 8}{\bar
h \over  < \bar H>},
\end{equation}
for which the solution $\hat U(h, \bar h)$ is given by:
\begin{equation}
\hat U(h,\bar h)=-{1 \over 8}{h^2  \over < H>}+C_1 h -{1 \over
8}{\bar h^2  \over <\bar  H>}+ \bar C_1\bar h.
\end{equation}

b)The second solution reads as:
\begin{equation}
W(h,\bar h)=-{1 \over {4(n+1)!}}{1 \over { (1+4J)}} \int dx < H>-
{1 \over {4(n+1)!}}{1 \over {(1+4 \bar J)}} \int dx <\bar  H>,
\end{equation}
Similarly to (a), the constraints (2.7) and (2.9) lead to the
differential equation:
\begin{equation}
\hat U-h{\partial \hat U \over \partial h}-\bar h{\partial\hat  U
\over \partial \bar h}={1 \over 4}{h^{1/2}  <H>^{1/2}}+{1 \over
4}{\bar h^{1/2}  <  \bar H>^{1/2}},
\end{equation}
which admits the solution:
\begin{equation}
\hat U(h,\bar h)={1 \over 2}h^{1/2}< H>^{1/2}+C_2 h+{1 \over
2}\bar h^{1/2}< \bar H>^{1/2}+\bar C_2 \bar h,
\end{equation}
where $C_1$ and $C_2$ are  complex constants.

Thus, like for HZ potential (2.11), we remark that the effective
potentials (2.19) and (2.22) provide new ways to parameterize the
glueballs interactions. The fact that there is no unique form of
the glueballs potential is due to the complexity of the QCD vacuum
structure which is encoded in the low energy theorems (2.8).

\section{The axion/dilaton Potentials}

\qquad As we have mentioned in the introduction of this paper, the
dilaton and the axion are scalar fields predicted by string
theory. In the language of the 4D gauge theory,  the axion usually
couples to the topological term while the dilaton couples to the
field strength term. In this section, we shall exploit these
features to build an effective potential in which we combine both
of them. In this regard, we shall extend the lagrangian of VVW
model \cite{VV,W} by including the dilaton degrees of freedom,
\begin{equation}
L_{VVW}=L_0(U,U^+)+L(A_\mu ,\phi , \theta )+{1 \over 8}i
\epsilon_{\mu  \nu \rho \sigma }G^{\mu  \nu \rho \sigma }Tr[ log
U+log U^+] +{f_\pi  \over 2 \sqrt {2}}Tr[  MU+MU^+],
\end{equation}
This lagrangian is established at large $N_c$ limit in which QCD
exhibits both confinement and chiral symmetry breaking.
$L_0(U,U^+)$ is a sigma  model lagrangian which describes the low
energy dynamics of the pseudoscalar mesons parameterized by the
$U$ complex matrix:
\begin{equation}
L_0(U,U^+)={1 \over 2} Tr(\partial _\mu U \partial ^\mu U^+).
\end{equation}
The $\theta$-term in the QCD fundamental lagrangian will be
substituted by $-{\theta \over 4} \epsilon_{\mu  \nu \rho \sigma
}G^{\mu  \nu \rho\sigma } $,
 where $G^{\mu  \nu \rho\sigma }$ is the field strength for a three-index field
 $A_{ \mu \nu \rho }$  given by \cite{VV,SHI}:
\begin{equation}
G_{\mu  \nu \rho \sigma }=\partial _\mu A_{ \nu \rho \sigma
}-\partial _\sigma  A_{ \mu \nu \rho }+\partial _\rho A_{  \sigma
\mu \nu }-\partial _\nu A_{  \rho  \sigma \mu };
\end{equation}
and where $A_{ \mu \nu \rho }$ is related to the gauge field $A_{
\mu }$  by the following expression:
\begin{equation}
 A_{ \nu \rho \sigma }={g^2 \over 96 \pi ^2 }[-A_\nu ^a
\overleftarrow{\partial} _\rho A_ \sigma ^a+A_\nu ^a
\partial _\rho A_ \sigma ^a+A_\rho ^a \overleftarrow{\partial} _\nu A_ \sigma ^a-A_\rho ^a \partial _\nu A_ \sigma
^a+A_\nu ^a \overleftarrow{\partial} _\sigma A_ \rho ^a -A_\nu ^a
\partial _\sigma A_ \rho ^a+2f_{abc}A_\nu ^a  A_ \rho ^bA_ \sigma ^c]
\end{equation}
$f_{abc}$ are the $SU(N_c)$ structure constants.\\ Since the
massive dilaton couples to the YM term $G_{\mu \nu }G^{ \mu \nu }$
as follows:
\begin{equation}
{\cal L}=-{1\over 4F(\phi )}G_{\mu \nu}G^{\mu \nu}-{1\over
2}\partial_ \mu \phi \partial^ \mu  \phi-{1\over 2}m^2
\phi^2+J_\mu^aA_a^\mu,
\end{equation}
we suppose that the kinetic term $ -dG_ {\mu \nu \rho \sigma
}G^{\mu \nu \rho \sigma }$ of the field $ A_{ \nu \rho \sigma }$
in  VVW model couples to the dilaton field via a given $ \phi $
function $1 \over F(\phi)$. The form of $ F(\phi)$ will be
discussed in the next section. Then we write the second term in
Eq.(3.1), which contains the gauge field $A_ \mu $, the dilaton
$\phi $ and the $\theta $-vacuum angle as :
\begin{equation}
L(A_\mu, \phi, \theta)={-d \over F(\phi)}G_ {\mu  \nu \rho \sigma
}G^{\mu  \nu \rho \sigma }-{\theta \over 4} \epsilon_{\mu  \nu
\rho \sigma }G^{\mu  \nu \rho \sigma }-{1 \over 2} \partial ^
\mu\phi
\partial_ \mu\phi-{1 \over 2}m^2 \phi^2,
\end{equation}
where $d$ is a positive constant and $m$ is the dilaton mass. \\
By considering the field $q(x)={1 \over 4} \epsilon_{\mu  \nu \rho
\sigma }G^{\mu  \nu \rho \sigma }$, the effective lagrangian at
large-$N_c$ reads as:
\begin{eqnarray}
L_{VVW}(U,U^+,q,\phi)&=& L_0(U,U^+)+{1 \over F(\phi)}N_c
q^2(x)-\theta \,q(x) +{1 \over 2}i\,q(x)\,Tr[
logU+logU^+]\\\nonumber && +{f_\pi \over 2 \sqrt {2}}\,Tr[
MU+MU^+] -{1\over 2}\partial_\mu \phi \partial^\mu \phi-{1 \over
2}m^2 \phi^2
\end{eqnarray}
Moreover, if The equation of motion of q(x) is used the expression
(3.7) reduces to:
\begin{eqnarray}
L(U, U^+ , \phi)&=&L_0(U,U^+)+{c\, F(\phi) \over 4N_c} \{\theta
-{1\over 2}\,Tr[ log U+log U^+]\} ^2\\ \nonumber &&+{f_\pi \over 2
\sqrt {2}}\,Tr[ MU+MU^+]- {1\over 2}\partial_\mu \phi
\partial^\mu \phi-{1\over 2}m^2 \phi^2,
\end{eqnarray}
with $c={4! \over d}$. For pure YM theory at large-$N_c$ coupled
to the dilaton field, the effective potential $W$ containing
$\theta$ and $\phi$ derived from the potential part of the
lagrangian (3.8) is given by:
\begin{equation}
W(\theta, \phi)=W(\theta=0, \phi)-{c F(\phi) \over 4 N_c}\, \theta
^2-{1 \over 2}m^2 \phi^2.
\end{equation}
As usual, we adopt the guess that the missing orders in the ${1
\over N_c}$ expansion provide a cosine form for the potential
$W(\theta, \phi)$, namely,
\begin{equation}
W(\theta, \phi)={c F(\phi) \over 4}N_c\, cos({\theta \over
N_c})-{1 \over 2}m^2 \phi^2.
\end{equation}
Therefore, the axion potential derived via: $\theta \to {a\over
f_a}$ correspondence, can be written in terms of the dilatonic
vacuum expectation value as:
\begin{equation}
W(\theta, \phi)={c <F(\phi )> \over 4}N_c\, cos({\theta \over
N_c})-{1 \over 2}m^2< \phi ^2>.
\end{equation}
Such formula is more general: It describes the axion potential for
any kind of dilaton-gluon coupling $F(\phi )$. If we consider the
particular case  $F(\phi) ={\phi \over f}$, the dilaton mass and
$<\phi ^2>$ can be substituted in (3.11) by using the following
relations \cite{DF}:
\begin{eqnarray}
&&m^2f^2_\phi=- {2\beta (g) \over g}<0|G_{\mu \nu }G^{\mu \nu
}|0>,\nonumber \\ &&4<\phi ^2>=-{g \over 2 \beta (g) }{f_\phi ^3
\over f}-f_\phi ^2,
\end{eqnarray}

with $ -{2\beta (g) \over g}= ({11\over 3}N_c-{2\over
3}N_f){\alpha_s \over 2\pi}=b{\alpha_s \over 2\pi}$,   $f_\phi$ is
the dilaton decay constant and $f $ represents a mass scale. The
axion potential (3.11) then  becomes
\begin{equation}
W(a, <\phi>)=-{1\over 8}{2\beta (g) \over g}(1+{g\over 2\beta (g)
}{f_\phi\over f})<0|G_{\mu \nu }G^{\mu \nu }|0> + {c<F(\phi)>
\over 4}N_c\, cos({a \over f_a N_c}),
\end{equation}
taking the same form as the HZ one, with the gluon vacuum
condensate term expressed as an expansion at leading order in
${f_\phi\over f}$ ( ${f_\phi\over f}\ll 1$). As a byproduct, For a
vanishing vacuum angle, the formula (3.11) reduces to the
expression of the dilaton potential,
\begin{equation}
W( \phi)={c <F(\phi)> \over 4}N_c -{1 \over 2}m^2 \phi^2.
\end{equation}
\quad The effective potentials given by Eqs.(3.9) and (3.12)
generalize the VVW one which corresponds to the particular case
when $F(\phi)=1$, the coupling of the gauge field to a massless
dilaton. Consequently, the lagrangian $L(U, U^+, \phi)$ in (3.8)
allows us to reconsider the HZ model by inclusion of the dilaton
contribution.

\section{New Interquark Potential}

\qquad  In QCD framework,  quarks and gluons (in general all
colored  objects) appearing in the fundamental QCD lagrangian
cannot exist as separate objects and then are absent from the
physical spectrum: such phenomenon is explained by the
confinement. Confinement in gauge theories provides one of the
most challenging problems in theoretical physics. Various quark
confinement models rely on flux tube picture. The latter emerges
through the condensation of magnetic monopoles and explain the
linear rising potential between color sources. However, a deep
understanding of confinement mechanism in still lacking. Recently
it has been shown in \cite{D} that a string inspired coupling of a
dilaton $\phi$ to the 4d $ SU(N_c)$ gauge fields yields a
phenomenologically interesting interquark potential $V_D(r)$ with
a linear confining term. Extension of gauge field theories by
inclusion of dilatonic degrees of freedom has gained considerable
interest. Particularly, Dilatonic Maxwell and Yang Mills theories
which, under some assumptions, possess stable, finite energy
solutions \cite{CT}. Indeed, the derivation performed in \cite{D}
is remarkable since it provides a challenge to monopole
condensation as a new quark confinement scenario. Therefore, our
objective in this section is to dedicate more efforts to the
investigation of this confinement generating mechanism with the
aim to derive a new interquark potential, extending Dick one.
\\
 To this end, we propose an  effective field theory described by the lagrangian,
\begin{equation}
L=-{1\over 4F(\phi )}G_{\mu \nu}G^{\mu \nu}-{1\over 2}\partial_
\mu \phi \partial^ \mu  \phi-{1\over 2}m^2 \phi^2+J_\mu^aA_a^\mu,
\end{equation}
where the coupling $-{1\over 4 F(\phi )}$  is function of the
dilaton field and m denotes the dilatonic mass. Such form appears
in several theoretical frameworks, for instance: ${1\over F(\phi
)}=e^{\phi \over {f_\phi }}$ as in string theory, ${1\over F(\phi
)}={\phi \over {f_\phi }}$ similar to the Cornwall-Soni model in
which ${1\over F(\phi )}$  parameterizes the coupling of gluon to
the glueball field \cite{GS}; while in Dick model, $F(\phi )$ is
given by: ${1\over F(\phi )}={\phi ^2 \over {f^2+\kappa \phi
^2}}$.\\ The analysis of the Coulomb problem of the dilatonic
theory in (4.1) is performed by considering a point like static
source described by the current density $J^ \nu _a=g \delta (r)
C_a \eta ^\mu$ where $C_a$ is the expectation value of the
$SU(N_c)$ generators for a normalized spinor  in  the color space
, satisfying the algebra identity $\Sigma _a  C_a^2={N_c^2-1 \over
{2 N_c}}$.\\ The equations of motion corresponding to $\phi$ and
$A_\mu$, inherited from the lagrangian (4.1) and emerging from
this source, read as,
\begin{eqnarray}
&&\partial ^2 \phi -m^2\phi = {1 \over {4}}{d(1/F(\phi )) \over d
\phi }G_{\mu \nu}G^{\mu \nu}\\ &&\partial_{\mu} ({1 \over F(\phi
)}G^{\mu \nu }_a)+g{1 \over F( \phi )}A_ \mu  ^b f_{ab}^c G^ { \mu
\nu  }_c = -J^ \nu _a ,
\end{eqnarray}
which yields
\begin{equation}
{d^2 \over dr^2}(r\phi)-m^2(r \phi)={\mu ^2 \over {2
r^3}}{d(F(\phi)) \over {d\phi}},
\end{equation}\\
where the abbreviation $\mu ={g \over 4 \pi}\sqrt {N_c^2-1 \over 2
N_c}$ is used. The equation (4.4) may be solved for a given
dilaton-gluon coupling $F(\phi)$ for which the interquark
potential $V(r)$ is expressed by the important formula \cite{CMS}:
\begin{equation}
V(r)=-{g\over {4\pi }}C \int {F(\phi(r)) \over r^2}dr.
\end{equation}

Such form of potential is very attractive since it extends the
usual Coulomb formula $V_c(r)\sim {1\over r}$, recovered from
(4.5) by taking $F(\phi )=1$. Moreover, for $F(\phi )=r^n$ with
$n\geq 2$, the formula (4.5) yields a confining potential. On the
other hand Eq.(4.5) may be also used to relate nonperturbative
effects such as QCD vacuum condensates to dilaton parameters $(m,
f)$ via comparison with known interquark potentials exhibiting QCD
power corrections, such as BHS potential \cite{BHS}.\\

For Dick model $V_D(r)$ which corresponds to the coupling $F(\phi
)=f^2+{\kappa \over \phi^2}$, the confining potential is given by
\begin{equation}
V_D(r)={{N_c^2-1}\over 2N_c}\{{\kappa g^2 \over 4\pi }{1 \over
r}-g f \sqrt {N_c \over 2(N_c-1)} log (e^{2mr}-1+my_0 )\},
\end{equation}
Thus $V_D(r)$ exhibits two behaviors which characterize the quark
interactions, namely: Besides Coulombian term which dominates at
short distances, the potential develops, at large distances, a
linear rising part, $V(r) = \sigma r$, describing the confining
phase,  where the string tension $\sigma$ is proportional to
dilaton parameters (m, f) as $\sigma \sim 2gfmr$. Recently, Dick
potential was successfully checked in \cite{CB} where it has been
shown that, if the dilaton is assigned a mass about $57 MeV$, the
spectrum of the charmonium and bottomnium are reproduced well.
Furthermore, the spin averaged energy levels of $B_c$ and $B^*_c$
systems have been estimated with values in good agreement with
other theoretical framework predictions.
\\
However, If we invoke the picture of the old string theory,
confinement is connected to the formation of a string between two
quarks (or quark-antiquark) separated by a distance $r$. If $r$
increases the string tension increases too. Then, for large
distances the string may be broken and colored fields may appear
as separate objects. In this regards, it is convenient to
parameterize  the confinement in another way which preserves color
singlet feature of the system. To this end, we shall derive a more
general confining potential than Dick one, where high powers in
$r$ may appear. We rewrite the equation of motion (4.4) as
follows:
\begin{equation}
y''-m^2 y={\mu ^2 \over {2 r^3}}{d(F(\phi)) \over {d\phi}},
\end{equation}
which, for Dick choice, reduces to:
\begin{equation}
y''-m^2y=-{\mu^2\over{ y^3}}.
\end{equation}
with $ y=r \phi$ and $y''={d^2y \over {dr^2}}$.

Note that Eq.(4.7) can be interpreted as corresponding to a
mechanical system with the action,

\begin{equation}
S= \int dr L_D= {1\over 2} \int dr [(y')^2 + m^2 y^2 + {\mu^2\over
{r^2}}F(y/r)] \qquad
\end{equation}

Consequently, the effective coupling $F(\phi)$ in (4.1) appears as
a part of interacting potential of a 1d quantum field theory.  In
this case, looking for solutions of (4.8) is similar to solving
the following equation of motion  \cite{CMS},

\begin{equation}
y''{\partial L_D\over {\partial y'}}+y'{\partial L_D\over
{\partial y}}+{\partial _r}^{exp} L_D=0
\end{equation}

Now in order to derive a potential with higher order in $r$,
instead of the equation (4.7), let's consider,
\begin{equation}
y''-m^2y=\mu^2 (b_0+b_1r+b_2r^2)
\end{equation}
where the  parameters $b_0$, $b_1$ and $b_2 $ are arbitrary and
are connected to the nonperturbative effects parameterized by the
quark and gluon vacuum condensates, as it will be shown later. The
explicit solution $\phi (r)=y/r$ of (4.11) is
\begin{equation}
\phi= {1 \over r}\{A e^{-mr}-{{\mu^2}\over {m^2}}\lbrace
(b_0+{{2b_2}\over m^2})+b_1r+b_2r^2 \rbrace \}
\end{equation}
where A is an integration constant. Substitution of Eq.(4.12) in
(4.7) leads to the following expression of $F(\phi(r))$ as a
function of $r$:
\begin{eqnarray}
F(\phi(r))&=&K-{2{\mu^2}\over {m^2}}\lbrace -(b_0+{{2b_2}\over
m^2}){b_0\over 2}r^2
 -(b_0+{{2b_2}\over m^2}){b_1\over 3}r^3-{{b_2^2}\over {2m^2}}r^4 +{{b_1b_2}\over {5}}r^5+{{b_2^2}\over {6}}r^6 \rbrace \nonumber \\
 &&-\lbrace  (3b_0+{{8b_1}\over m}+{{30b_2}\over m^2}){1\over m^2}+(3b_0+{{8b_1}\over m}+{{30b_2}\over m^2}){r \over m}\nonumber \\
&& +(b_0+{{4b_1}\over m}+{{15b_2}\over
m^2}){r^2}+(b_1+{{5b_2}\over m}){r^3}+b_2r^4  \rbrace Ae^{-mr}.
\end{eqnarray}

Finally, by invoking the formula (4.5), we can easily obtain the
interquark potential $V(r)$:

\begin{eqnarray}
V(r)&=&{\alpha K\over r}-{2{\alpha \mu^2}\over {m^2}}\lbrace
-(b_0+{{2b_2}\over m^2}){b_0\over 2}r
 -(b_0+{{2b_2}\over m^2}){b_1\over 6}r^2-{{b_2^2}\over {m^2}}{r^3\over 6} +{{b_1b_2}\over {20}}r^4+{{b_2^2}\over {30}}r^5 \rbrace \nonumber \\
 &&-\alpha Ae^{-mr}\lbrace  (3b_0+{{8b_1}\over m}+{{30b_2}\over m^2}){1\over { m^2r}}+(b_0+{{5b_1}\over m}+{{22b_2}\over m^2}){1 \over m}\nonumber \\
&& +(b_1+{7b_2\over m}){r\over m}+{b_2\over m}r^2  \rbrace.
\end{eqnarray}
where $\alpha={g \over 4\pi}C$. This form of potential containing
higher power of $r$ is very interesting thanks to the presence of
several terms which come with different signs and have not the
same behaviours: Indeed, since some terms provide a confinement
effect while  the others have a deconfinement effect, the total
interquark potential should not diverge for large $r$. This
potential can be compared with the Bian-Huang-Shen potential
$V_{BHS}$ calculated in terms of the QCD vacuum condensates in the
QCD background fields \cite{BHS}
\begin{equation}
V_{BHS}(r)=(-{4 \over {3\pi } }g^2+B_f){1 \over {
r}}+Dr+Fr^3+Gr^5+ (-{B_f \over r}+C_fr+F_fr^2)e^{-m_fr}
\end{equation}
where the parameters $B_f$, $C_f$, $D$, $E_f$, $F$ and $G$ are
expressed through the following condensates (up to dimension 6)
$<0|\bar \psi \psi|0>$,$ <0|{\alpha \over \pi}G^2|0>$,$ <0|\bar
\psi \sigma G \psi|0>$, $<0|\bar \psi \Gamma _1 \psi \bar \psi\
\Gamma_ 2 \psi|0>$   and    $ f_{abc}<0|G_a^{\mu \nu}G^b_{\nu
\rho}G^\rho _{\mu c}|0>$. Moreover, If we assume that terms with
an exponential behaviour have a negligible contribution, in both
potentials $V(r)$ and $V_{BHS}(r)$ (this corresponds to the
behaviour of the potentials at large distances) with $b_1$ sets to
zero in Eq.(4.14), we will see that only terms with odd powers in
$r$ survive. This result is in perfect agreement with the BHS
hypothesis suggesting that for higher dimension condensates, the
interquark potential can be written as:
\begin{equation}
V(r)=-{\alpha \over { r}}+\Sigma_n(-1)^nC_{2n+1}r^{2n+1}.
\end{equation}
depending only on odd powers of $r$ with alternating signs and
where the coefficients $C_{2n+1}$ are function of the QCD
condensates. At the same time, the confrontation of $V(r)$ with
$V_{BHS}(r)$ (resp. $V_D(r)$ generates several relations between
$(b_0,b_1,b_2)$ (resp. $(f,m)$) and the vacuum condensates which
may provide an important piece of information in the search of the
dilaton.

\section{Conclusion}

\qquad In this paper we have reviewed the derivation procedure of
the axion potential as given by HZ and have proposed the
alternative forms for $U(h,\bar h)$ gluodynamics effective
potentials. We have seen that for the HZ solution, we can obtain
the axion potential as it is written in Eq.(2.13) directly  by
setting the glueball field $h$ in Eq.(2.16) as: $ h={E }e^{iD' }$.
As a by-product, we have found that the physical field $\eta$  is
expressed in terms of the $\theta$ vacuum and the pseudo-Goldstone
fields given by the $U$ matrix. The axion potential in (2.15),
which is parameterized by the vacuum condensates, has the same
form as a class of inflationary potentials describing the vacuum
potential, generally deduced by invoking the inflaton field
\cite{KM}. Next, we have extended the VVW model by the
implementation of the dilaton field. The new VVW-lagrangian allows
us to derive the dilaton potential and to regenerate the axion one
which take into account dilatonìc degrees of freedom. On the other
hand, as the QCD vacuum has a non trivial structure and the
physical spectrum does not contain colored objects, many potential
models have been suggested to describe the confining phases
characterizing the quark-quark interactions. We have derived in
(4.14) a new interquark potential containing terms with high power
of  $r$. The latter emerges from a string inspired model with an
effective dilaton-gluon coupling. The good agreement between V(r)
and $V_{BHS}(r)$, in the particular case of $ b_1=0$, allows us to
adopt the general expression (4.16) with only odd powers of $r$
and alternating signs as an adequate confining potential form for
higher dimension vacuum condensates.\\

{\bf Acknowledgments}\\
We are grateful to Prof. J. da Providencia
for his collaboration in the early stage of this work. We also
wish to thank Prof. T. Lhallabi for discussions.\\ This work is
supported by the program PARS/ 27.372/98/CNR and the CNPRST/ICCTI
convention 340/00 CNR.

\end{document}